\documentclass[twocolumn,aps,showpacs,floatfix,superscriptaddress]{revtex4}
\pdfoutput=1 
\usepackage{amsmath,amssymb,eucal,graphicx}

\begin{document}
\title{Molecular Spiders on the Plane}
\author{Tibor Antal}
\affiliation{School of Mathematics, Edinburgh University, Edinburgh, EH9 3JZ, UK}
\author{P. L. Krapivsky}
\affiliation{Department of Physics, Boston University, Boston, MA 02215, USA}

\begin{abstract}
Synthetic bio-molecular spiders with  ``legs'' made of  single-stranded segments of DNA can move on a surface covered by single-stranded segments of DNA called substrates when the substrate DNA is complementary to the leg DNA. If the motion of a spider does not affect the substrates, the spider behaves asymptotically as a random walk. We study the diffusion coefficient and the number of visited sites for spiders moving on the square lattice with a substrate in each lattice site. The spider's legs hop to nearest-neighbor sites with the constraint that the distance between any two legs cannot exceed a maximal span. We establish analytic results for bipedal spiders, and investigate multi-leg spiders numerically. In experimental realizations legs usually convert substrates into products (visited sites). The binding of legs to products is weaker, so the hopping rate from the substrates is smaller. This makes the problem non-Markovian and we investigate it numerically. We demonstrate the emergence of a counter-intuitive behavior --- the more spiders are slowed down on unvisited sites, the more motile they become. 
\end{abstract}

\maketitle

\section{Introduction}

Recent advances in DNA nano-fabrication technology (see \cite{milan,seeman,seeman2,pierce}) have led to the constructions of multi-pedal walking molecular devices. The first was a bipedal object \cite{seeman2} walking on a one-dimensional path by DNA set strands with nucleic acid domains complementary to molecular imprints on the device legs and the substrate. Since then, several other similar bipedal DNA walkers have been synthesized (see \cite{pierce,bath,seeman3,transporter} and a review \cite{synt_motors}). 

A different molecular design has been implemented in \cite{exp1,exp2}. The resulting objects, known as molecular spiders, usually have many legs. Each leg (a short single strand of DNA) can bind to the substrate through the Watson-Crick base pair formation. A bound leg can either detach from the substrate without modifying it, or it can catalyze the cleavage of the substrate creating two product strands. The lower product remains bound to the surface, while the upper product is free to float away in solution. There is a residual binding of the leg to the remaining product, but it is weaker than the leg-substrate binding. Utilizing the effect of the spider's motion on the molecular tracks it is possible to design environments where molecular spiders demonstrate some basic robotic behaviors \cite{exp2}. 

When a spider is released on a surface coated with oligonucleotide substrates, it can cleave thousands of substrates before eventually detaching. The small size of spiders makes experimental observation of their motion very challenging. Atomic force microscopy imaging and single-molecule fluorescence studies have been successful to a certain degree \cite{exp2}, yet neither the details of the spider's gait nor the individual paths of spiders have been resolved with sufficient certainty. Perspectives and challenges of the experimental work are surveyed in \cite{milan2}. 

A number of modeling studies describing the motion of molecular spiders have been recently carried out. The motion of a single spider on a one-dimensional track has been investigated in Refs.~\cite{spider1d,memo}. The first article \cite{spider1d} ignores the difference between the substrate and the product and makes a number of simplifying assumptions about the gait of the spiders. There are no limitations on the number of legs, however. The chief result of Ref.~\cite{spider1d} is that the spider (which is a complicated self-interacting multi-leg object) can be replaced by a particle characterized by a single number, the diffusion coefficient; for the simplest gaits, the diffusion coefficient was analytically computed. In \cite{memo} we mainly considered bipedal spiders, but took into account that spiders affect the substrate (turn it into products). Despite the non-Markovian nature of the problem, the coarse-grained behavior turned out to be surprisingly simple, namely the difference in residence times on the substrate and the product leads to the effective bias into the unvisited region. 

Recent papers \cite{zucker,zucker2,darko,darko2} utilized more detailed and complicated models mimicking the gait of spiders moving on one-dimensional tracks, the possibility of the detachment, etc. These studies numerically confirm the tendency of spiders to move into the unvisited region leaving behind the trail of the product. This key feature was observed experimentally \cite{exp1,exp2} and proved theoretically \cite{memo} in the realm of simple models. An interesting new feature noticed in \cite{darko} is the emergence of a super-diffusive growth of the mean-square displacement, $\langle x^2(t)\rangle \sim t^\alpha$ with $1<\alpha<2$, which holds on a surprisingly long time span; eventually, the super-diffusive growth crosses over to the diffusive growth. Several  rigorous results concerning the asymptotic behaviors (limit theorems, transience, recurrence, and rate of escape) of molecular spiders have been established in \cite{bipeds,Gal1}. In Refs.~\cite{Gal2,juhasz} the motion of spiders in random environments has been studied. 

\begin{figure}
\centering
\includegraphics[scale=0.13]{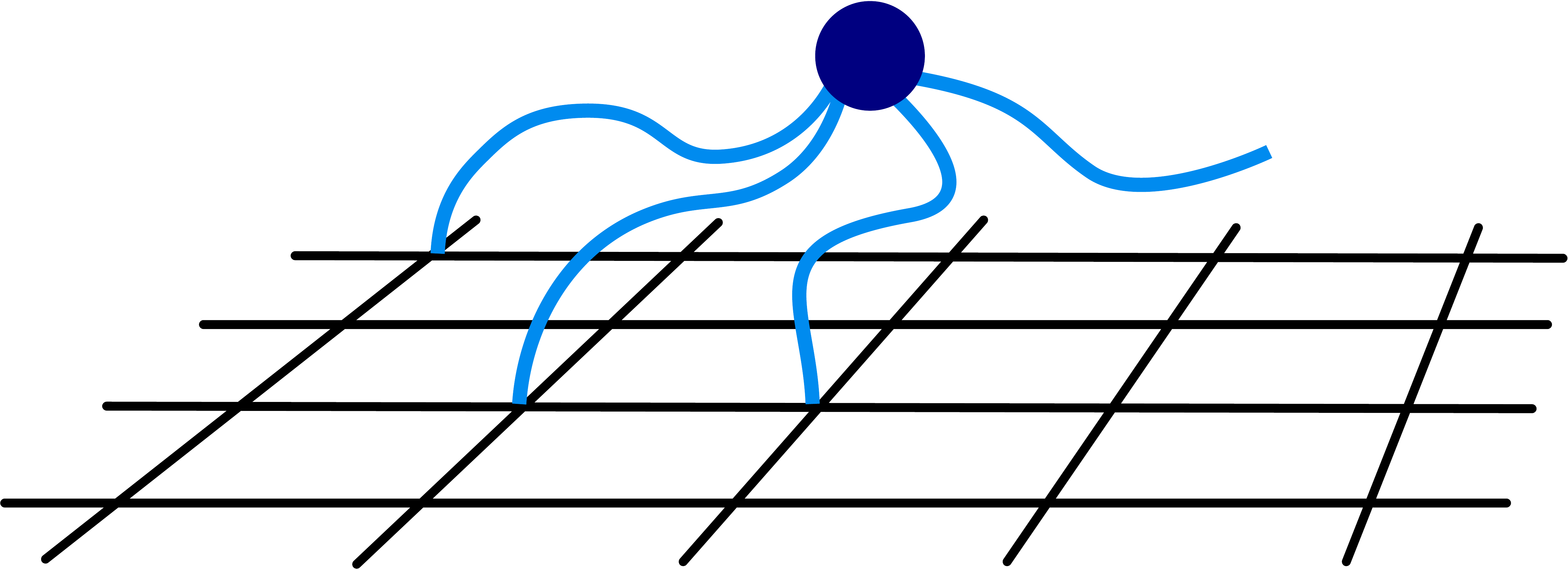}
\caption{A molecular spider with four legs moving on the square lattice. The distances between any two legs should not exceed a threshold value (the maximal span $S$). As long as this constraint is obeyed, each leg can hop to the empty (not occupied by another leg) nearest neighboring site. As long as the rate of leg attachment greatly exceeds the rate of leg detachment, all legs are attached most of the time and for relatively short time intervals one of the leg is detached as illustrated above.}
\label{twodim_ill}
\end{figure}

As in our previous work \cite{spider1d,memo}, throughout this paper we will assume an idealized gait --- the goal is not to mimic the complicated (and poorly known) gait of molecular spiders, but to qualitatively understand spiders' macroscopic characteristics in the realm of simple models \cite{models}. Previous theoretical analyses \cite{spider1d,memo,zucker,zucker2,darko,darko2} have been focused on the motion of molecular spiders on one-dimensional tracks, while the goal of this work is to study a single spider moving on a two-dimensional lattice. If not stated otherwise, we tacitly assume that the rate of attachment greatly exceeds the rate of detachment. In this case, the relative time when one leg is detached (this situation is illustrated on Fig.~\ref{twodim_ill}) is negligible and hence the possibility that two or more legs are detached simultaneously can be disregarded. 

The following properties of molecular spiders will {\em always} be assumed:
\begin{itemize}
\item {\bf Hopping}: When a leg detaches, it re-attaches to a {\em neighboring} site.
\item {\bf Exclusion}: Two legs cannot be attached to the same site. 
\item {\bf Constraint}: The distance between {\em any} two legs does not exceed a certain maximal span. 
\end{itemize}
The restriction to the nearest neighbor hopping can be relaxed (long-distance gaits have been probed in the one-dimensional setting \cite{spider1d}); the exclusion is of course the fundamental feature. The last property is the simplest constraint that assures the compactness of the spider. For one-dimensional spiders several types of constraint were considered, and the diffusion coefficients of these spiders were exactly obtained in most cases, due to mappings to exclusion processes \cite{spider1d}. 

The two-dimensional case is particularly important in current experiments. The actual situation is rather complicated, e.g., in some experiments there are a few layers of the substrate and hence a quasi two-dimensional setting seems more appropriate; additionally, the substrates do not form a perfect square lattice. Nevertheless, we shall assume that a spider with aforementioned simple gait is placed on a square lattice \cite{other}. We emphasize that the constraint regarding the separations between the legs roughly describes real molecular spiders. One realization of the constraint which is convenient for the numerical implementation defines the distance between points $(x_1,y_1)$ and $(x_2,y_2)$ via $\max(|x_1 - x_2|, |y_1 - y_2|)$. The neighborhood in this metric is geometrically a square, and it is often called the von Neumann neighborhood. For example, the simplest ``von Neumann'' spider is the bipedal spider with legs separated by distance $S=1$ at most.  There are four possible configurations: horizontal, vertical, and two diagonal 
\begin{equation}
\label{simpconf}
\begin{array}{cc} \circ&\circ\\ \bullet&\bullet \end{array} ~,~~~
\begin{array}{cc} \bullet&\circ\\ \bullet&\circ \end{array} ~,~~~
\begin{array}{cc} \circ&\bullet \\ \bullet&\circ \end{array} ~,~~~
\begin{array}{cc} \bullet&\circ \\ \circ&\bullet \end{array} 
\end{equation}
where $\bullet$ represents a leg, and $\circ$ an empty site. We found that the diffusion coefficient \cite{D_convention} of this spider is equal to $1/4$. Generally von Neumann spiders are more amenable to analysis, and we study them as well as more realistic Euclidean spiders. 

The rest of this paper is organized as follows. Section \ref{von} is devoted to von Neumann spiders. In Sec.~\ref{bi} we present main results for bipedal Euclidean spiders; the detailed derivations are given in the Appendix.  An analysis of molecular spiders simplifies when the maximal span increases and in Sec.~\ref{manyleg} we describe corresponding asymptotic behaviors. In the following sections we relax some of the assumptions about the spider gait, the influence of the spider's motion on the environment, etc. In Sec.~\ref{memory} we consider the influence of memory. We model the difference between the product and the substrate by postulating that the leg spends (on average) more time at newly visited sites, i.e., on the substrates. This slowdown in comparison with the motion on the products leads to faster (covering more unvisited sites) spiders; the reason for this phenomenon is an effective bias towards unvisited sites. This behavior has been observed and explained in one dimension \cite{memo}, and it continues to hold in two dimensions.  In Sec.~\ref{binding} we investigate what happens when the attachment rate is finite. We compute the mean time the spider spends on the surface and show that the probability to remain attached decays exponentially if the attachment rate greatly exceeds the detachment rate. We summarize our findings in Sec.~\ref{disc}.

\section{Von Neumann spiders}
\label{von}

A spider is quantified by lattice points ${\bf r}_i=(x_i,y_i)$ with $i=1,\dots,L$ which describe the positions of its legs. We assume that the spider has maximal span $S$, so the distance between any two legs is $\leq S$. Each leg, when allowed, hops to neighboring sites (up, down, left, or right) at rate 1 in each direction.  In this section we use the metric which assigns the distance between any two legs ${\bf r}_i$ and ${\bf r}_j$ according to the rule
\begin{equation*}
|{\bf r}_i-{\bf r}_j|_\infty=\max(|x_i - x_j|, |y_i - y_j|)
\end{equation*}
In this metric, the neighborhood around the origin, i.e. the disk $|{\bf r}|_\infty\leq S$, is the square in Euclidean metric [see the example in \eqref{simpconf}]; such a neighborhood is often called the von Neumann neighborhood.

A spider with maximal span $S$ between the legs should therefore occupy the square with $(S+1)^2$ lattice sites; after shifting, this square becomes
\begin{equation}
\label{square}
\{(x,y): x=0,\ldots,S; y=0,\ldots,S\}
\end{equation}
This spider can therefore have at most $(S+1)^2$ legs. Only spiders with
\begin{equation}
\label{maxL}
L\leq S(S+1)
\end{equation}
legs are mobile. More precisely, spiders with more legs than the above upper limit, 
$S(S+1)<L<(S+1)^2$, have a few legs which can move, but each such spider forever remains within its surrounding  square \eqref{square}, provided that we ignore multiple legs being detached simultaneously.

Let us first calculate the total number of possible configurations of the legs. To avoid multiple counting of configurations which are obtained by translation, we use the convention that there must be a leg both in the bottom row ($y=0$) and in the leftmost column ($x=0$), as in the example in \eqref{simpconf}. With this convention the total number of configurations is
\begin{equation}
\label{number-conf}
\mathcal{C}(L,S)=\binom{(S+1)^2}{L}-2\binom{S(S+1)}{L}+\binom{S^2}{L}
\end{equation}
To establish \eqref{number-conf} let us ignore for a moment the aforementioned convention restricting the position of one leg. The number of such unrestricted configurations of a spider with $L$ legs inside the $m\times n$ rectangular is equal to $\Omega(m,n) = \binom{mn}{L}$. To obtain $\mathcal{C}(L,S)$ we take the number of configurations  without restriction $\Omega(S+1,S+1)$, subtract the number of configurations without a leg in the bottom row $\Omega(S+1,S)$ or in the leftmost column $\Omega(S,S+1)$, and then add the number of configurations without a leg in both the bottom row and the leftmost column $\Omega(S,S)$, since we subtracted it twice. This leads to \eqref{number-conf}. 

We are interested in the properties of the center of mass of the legs
\begin{equation}
\label{centermass}
{\bf R} = \frac{1}{L}\sum_{j=1}^L  {\bf r}_j 
\end{equation}
Due to the symmetry of the hopping rules, the mean position of the spider does not change $\langle {\bf R}\rangle=0$, and we are interested in the variance $\langle {\bf R^2}\rangle$ of the center of mass.

\subsection{Bipedal spiders}

Here we show that the bipedal spider with arbitrary maximal span $S$ has the diffusion coefficient 
\begin{equation}
\label{D-bi-S}
 D = \frac{1}{2}\left( 1 -\frac{1}{2S} \right)
\end{equation}
The bipedal spider with $S=1$ corresponds to the bipedal spider with $\ell=\sqrt{2}$ in the Euclidean version, and from \eqref{D-bi-S} we indeed recover the already known result $D=1/4$.  

To derive \eqref{D-bi-S} we start by noticing that bipedal spiders are ``completely symmetric": In each configuration the spider hops at the same rate in each direction. For completely symmetric spiders on the square lattice, a remarkably simple formula 
\begin{equation}
  \label{compD} 
  \langle {\bf R}^2\rangle =\frac{\omega t}{L^2}=4Dt, \quad  \mathrm{with}\quad D=\frac{\omega}{4 L^2}
\end{equation}
for the mean-square of the center of mass, and respectively for the diffusion coefficient of the spider, was established in \cite{spider1d}. In Eq.~\eqref{compD} we denote by $\omega$ the total rate the spider hops which is averaged over all stationary states. In general, for spiders with any number of legs and with symmetric hopping rates, the transitions between any two connected configurations occur at the same rates and hence all configurations have the same stationary probability $1/{\cal C}$. Therefore the total rate which is averaged over all stationary states, in short the stationary average rate, is equal to
\begin{equation}
\label{averrate}
  \omega = \frac{1}{4\mathcal{C}}\sum_{j=1}^{\mathcal{C}} \omega_j
\end{equation}
where $\omega_j$ is the total  hopping rate from configuration $j$. Consequently, \eqref{compD} can be rewritten as
\begin{equation}
\label{compD2} 
  D= \frac{1}{4L^2\mathcal{C}}\sum_{j=1}^{\mathcal{C}} \omega_j
\end{equation}
(Note that for the symmetric random walk on the square lattice $L=1$ and $\omega=4$, hence the diffusion coefficient is one.) 

Let us first calculate the average hoping rate to a given arbitrary direction, say to the right. Notice that in most configurations both legs can jump to the right, but in some, only one of them. It is easier to enumerate the number of these later configurations, which is where one leg is blocked and cannot hop to the right. There is only one configuration where the two legs are next to each other in the same row $\bullet\bullet$, due to the convention that a leg is needed both in the bottom row and the leftmost column. In this configuration only the right leg can hop to the right. There are $S+1$ configurations where one leg is in the bottom left corner $(0,0)$, and the other is at maximal distance $x=S$, hence only the left leg can hop to the right. There are further $S$ configurations where the right leg cannot hop: one leg is in the bottom right corner $(S,0)$, and the other leg is in the leftmost column, and in row $y=1, \dots, S$. Hence all together there are $1+(S+1)+S=2(S+1)$ configurations where only one leg can hop to the right, and ${\cal C}-2(S+1)$ where both can. Now using the fact that the average rate is the same for all four directions, equation \eqref{averrate} leads to
\begin{equation}
\label{omega_S}
 \omega = \frac{2(S+1)+2[{\cal C}-2(S+1)]}{{\cal C}} = 2 - \frac{1}{S}
\end{equation}
where we have taken into account that the number of configurations for bipedal spiders is ${\cal C}=2S(S+1)$, as it follows from \eqref{number-conf}. Substituting \eqref{omega_S} into \eqref{compD} we arrive at the diffusion coefficient \eqref{D-bi-S}.

\subsection{Multipedal spiders}

It is much more challenging to compute the diffusion coefficient for multi-leg spiders. The chief reason is that spiders with more than two legs are not completely symmetric and hence one cannot use \eqref{compD}. To show the lack of symmetry it suffices to provide a configuration where the left and right hopping rates of the spider are different. Consider for example a tripod with two legs being in the same column and the third leg being at the maximal distance $S$ in the $x$ direction from both of the other legs, as illustrated here
\begin{equation}
\label{tripod} 
\begin{tabular}{|ccc|}
$\bullet$ & $\longleftarrow S \longrightarrow$ & $\bullet$\\
$\bullet$ &&
\end{tabular}
\end{equation}
This spider can hop to the right at rate two and to the left at rate one. (More precisely, both legs on the left can hop to the right, but not to the left; the leg on the right can hop to the left, but not to the right.) Similar configurations can be easily constructed for any spiders with $L\ge 3$ legs.

The matrix method described in Ref.~\cite{spider1d} can be generalized to arbitrary dimension and in principle it allows one to analytically determine the diffusion coefficient for any spiders with sufficiently small number of legs. However, even in the simplest examples the exact calculations are rather cumbersome. For instance, even for the simplest $S=2$ tripod on the square lattice the number of configurations is equal to 48 [see Eq.\eqref{number-conf}], so the computation of $D$ leads to the necessity to diagonalize a $48\times 48$ matrix; obtaining this matrix is very laborious. 

We performed simulations for the simplest spiders to measure their diffusion coefficients. This quantity is relatively easy to measure by probing the asymptotic of the mean-square displacement. It turns out that the  correction to the true asymptotic decays is rather small, namely,
\begin{equation}
\label{Dnomemcorr}
  \langle {\mathbf R}^2 \rangle \approx 4Dt \left(1 + \frac{a}{t} \right)
\end{equation}
for multi-pedal spiders. (Similar corrections were observed in one dimension.) Note that there are no corrections at all for continuous time random walks (the mono-pedal spiders). The measured values of $D$ are summarized in Table \ref{Dtab}. 

To detect the motion of a single spider is still experimentally impossible, while various techniques allow one to count the total number of visited sites \cite{exp1}.  Since numerous spiders are usually released \cite{exp1}, dividing the actually observed total number of visited sites by the number of spiders makes the {\em mean} number of sites visited by a spider accessible. In the limit when the density of spiders is very low, the same result will emerge if we take a single spider and average the number of different visited sites over many realizations. 

For a symmetric random walk which hops at rate one in each direction (so its diffusion coefficient is $D=1$), the mean number of different sites visited by the random walk scales as \cite{wijland,book}
\begin{equation}
\label{different-spider} 
 \langle N\rangle = \frac{A t}{\ln B t} + {\cal O}\! \left[\frac{t}{(\ln t)^3} \right]
\end{equation}
in the large time limit. The amplitudes are  $A=4 \pi$ and $B=32\exp(C_E-1)=20.97\dots$
where $C_E=0.577215\dots$ denotes Euler's constant. (In Ref.~\cite{wijland}, the random walk with diffusion coefficient $D=1/4$ has been analyzed; to recast the prediction of \cite{wijland} to our setting, where $D=1$, we rescaled time by a factor 4.) 

We simulated the motion of different spiders and we obtained the same asymptotic behavior \eqref{different-spider}, with $A$ and $B$ depending on the number of legs $L$, and the constraint $S$.
We are mainly interested in the leading order behavior $A$, however, fitting also the next correction $B$ is unavoidable in order to get an estimate for $A$, due to the large sub-leading corrections. The results for the coefficient $A$ are summarized in Table \ref{Atab}.

One can see from Tables \ref{Dtab} and \ref{Atab} that, according to simulations, $D(L,S)$ and $A(L,S)$ are monotonically decreasing functions of $L$, and monotonically increasing functions of $S$ for multi-pedal spiders. The simplest conjecture is that asymptotically the spider is indistinguishable from the random walk on the square lattice. Mathematically, this would imply that $A=4\pi D$. Simulation results for the diffusion coefficient show that $A$ is slightly different than that.

\begin{table}
\begin{tabular}{l || ccc}
 & $S=1$ & $S=2$ & $S=3$\\
\hline
$L=1$  &	1 & 1 & 1\\
$L=2$  &	0.25 & 0.375 & 0.417\\
$L=3$  &	  -   & 0.191   & 0.241\\
$L=4$  &	  -   & 0.0972 & 0.152\\
$L=5$  &	  -   & 0.0464 & 0.104\\
\end{tabular}
\caption{Simulation results for the diffusion coefficients $D$ for spiders with $L$ legs and constraint $S$. For the random walk ($L=1$) the diffusion coefficient is $D=1$; for bipedal spiders ($L=2$) we have exact results given by \eqref{D-bi-S}.}
\label{Dtab}
\end{table} 

\begin{table}
\begin{tabular}{l || ccc}
 & $S=1$ & $S=2$ & $S=3$\\
\hline
$L=1$  &	12.57 & 12.57 & 12.57\\
$L=2$  &	3.21   & 4.83   & 5.35\\
$L=3$  &	  -      & 2.38    & 3.20\\
$L=4$  &	  -      & 1.30    & 2.08\\
$L=5$  &	  -      & 0.627  & 1.44\\
\end{tabular}
\caption{Simulation results for the amplitude $A$ in the asymptotic law for the number of visited sites $\langle N \rangle = At/\ln t$. Spiders have $L$ legs and constraint $S$. For the random walk $L=1$ and the amplitude is $A=4\pi$.}
\label{Atab}
\end{table}

\section{Euclidean Spiders}
\label{bi}

In this section we consider the more realistic Euclidean spiders. Thus we use the standard Euclidean metric to measure the distance between the legs:
\begin{equation*}
 |{\bf r}_i-{\bf r}_j| = \sqrt{(x_i - x_j)^2+(y_i - y_j)^2}
\end{equation*}
The simplest Euclidean spider is bipedal with legs separated by maximal distance $\ell=\sqrt{2}$. This spider is identical to the simplest $L=2, S=1$ von Neumann spider; it is characterized by four configurations \eqref{simpconf}, and it has diffusion coefficient $D=1/4$. 

For the bipedal spider with maximal separation $\ell=2$, there are six different configurations --- four configurations \eqref{simpconf} and two additional configurations, the horizontal configuration $\bullet\circ\bullet$ and its vertical cousin. The computation of the diffusion coefficient gives $D=1/4$ [see Eq.~\eqref{D_6} in the Appendix], so this spider has the same diffusion coefficient as the previous one. This surprising result is a coincidence rather than a rule. We computed diffusion coefficients for bipedal spiders with many other maximal distances. For $\ell=\sqrt{5}$, four new configurations (with legs separated by the move of a knight in chess) appear. Next change occurs for $\ell=2\sqrt{2}$ when two new ``long'' diagonal configurations arise. Then for $\ell=3$, the horizontal configuration $\bullet\circ\circ\,\bullet$ and analogous vertical configuration become possible. Varying $\ell$ up to $\sqrt{50}$, the total number of distinct allowed configurations $\mathcal{C}$ and the diffusion coefficient exhibit the behaviors summarized in Table \ref{euctab}.

\begin{table}
\medskip 
\begin{tabular}{r||r|r|r|}
$\ell$&$|\mathcal{C}|$&$D$&$(1/2-D)\, \mathcal{C}$\\
\hline
\hline
$\sqrt{2}$ &4 &1/4   &1\\
2              &6 &1/4   &3/2\\
$\sqrt{5}$ &10&7/20  &3/2\\
$\sqrt{8}$ &12&3/8   &3/2\\
3              &14&5/14  &2\\
$\sqrt{10}$&18&7/18  &2\\
$\sqrt{13}$&22&9/22  &2\\
4               &24&19/48 &5/2\\
$\sqrt{17}$&28&23/56 &5/2\\
$\sqrt{18}$&30&5/12  &5/2\\
$\sqrt{20}$&34&29/68 &5/2\\
5               &40&17/40 &3\\
$\sqrt{26}$&44&19/44 &3\\
$\sqrt{29}$&48&7/16  &3\\
$\sqrt{32}$&50&11/25 &3\\
$\sqrt{34}$&54&4/9   &3\\
6               &56&49/112&7/2\\
$\sqrt{37}$&60&53/120&7/2\\
$\sqrt{40}$&64&57/128&7/2\\
$\sqrt{41}$&68&61/136&7/2\\
$\sqrt{45}$&72&65/144&7/2\\
7               &74&33/74 &4\\
$\sqrt{50}$&84&19/42 &4\\
$\gg 1$&$\approx \pi\ell^2/2$&$\approx 1/2-(\pi\ell)^{-1}$&$\approx \ell/2$\\
\hline
\end{tabular}
\caption{Total number of configurations and diffusion coefficients for Euclidean spiders.}
\label{euctab}
\end{table}

The diffusion coefficient tends to increase with $\ell$, yet its
behavior is somewhat erratic and it can occasionally decrease
($\frac{5}{14}<\frac{3}{8},\,\frac{19}{48}<\frac{9}{22},\,
\frac{17}{40}<\frac{29}{68},\,\frac{49}{112}<\frac{4}{9},\,\frac{33}{74}<\frac{65}{144}$,  etc.). The last
column reveals remarkable hidden regularities --- the quantity
$(1/2-D)\,|\mathcal{C}|$ is always half-integer with equilibrium patches of
increasing length punctuated by upward jumps by 1/2.  These intriguing
observations are explained by the neat general formula
\begin{equation}
\label{D-vs-L} 
D=\frac{1}{2}\left[1-\frac{\lfloor\ell\rfloor+ 1}{\mathcal{C}}\right]
\end{equation}
where $\lfloor\ell\rfloor$ is the integer part of $\ell$.  

The derivation of \eqref{D-vs-L} is somewhat lengthy (see the Appendix), but it is just an application of general formula \eqref{compD2} for the diffusion coefficient. The dependence of $\mathcal{C}$ and rates on $\ell$ is non-trivial and cannot be deduced analytically for an arbitrary $\ell$. Indeed, the problem of counting the total number of configurations is equivalent to the problem of computing $\mathcal{N}_\ell$ which gives the total number of lattice sites within the disk of radius $\ell$; more precisely, $\mathcal{C}=(\mathcal{N}_\ell-1)/2$. The investigation of $\mathcal{N}_\ell$ constitutes the celebrated Gauss problem \cite{guy}. Of course, $\mathcal{N}_\ell$ approximately grows as the area, $\mathcal{N}_\ell\approx \pi\ell^2$; the deviation from this dominant growth law is extremely difficult to probe analytically. (The sub-leading asymptotic is unknown. More precisely, the proven upper bound on the growth of the sub-leading asymptotic is substantially weaker than the conjectural one; proving the conjectural asymptotic is known to be equivalent to proving the Riemann conjecture.) This subtlety is irrelevant as long as we are satisfied with the leading asymptotic. In our problem the case of large $\ell$ is particularly simple as both legs essentially diffuse independently and therefore the diffusion coefficient of the center of mass is very close to the half of the diffusion coefficient of the random walk. Thus $D_\ell\to 1/2$ when $\ell\to\infty$. Note that $D<1/2$ for any two-leg spider. Indeed, since $\omega_j\leq 8$ (the equality occurs when each leg is allowed to hop to each of the four neighbors), the sum on the right-hand side of \eqref{compD2} cannot exceed $8\times \mathcal{C}$ which proves $D<1/2$.

The derivation of \eqref{D-vs-L} and a detailed description of the configurations for the threshold up to $\ell=\sqrt{50}$ are given in the Appendix.

\section{Spiders with large maximal span}
\label{manyleg}

Interactions between legs (due to exclusion and the maximal span constraint) imply that molecular spiders are complicated self-interacting objects. Therefore it is very difficult to compute the dependence of the diffusion coefficient $D(L,S)$ of a spider on the number of legs $L$ and the maximal span $S$. The behavior of $D(L,S)$ simplifies when the maximal span $S$ becomes large. In the $S\to \infty$ limit, typical separations between legs grow with time thereby making exclusion asymptotically negligible. Therefore in this limit we can treat legs as non-interacting random walkers. The diffusion coefficient of such spider is $D(L, S=\infty)=1/L$.  In this section we derive this result and then argue that the finite $S$ correction has the $1/S$ form.

\subsection{Non-interacting legs}

Consider a spider with $L$ non-interacting legs. Each leg performs a random walk with hopping rates one in each direction. The mean and mean-square displacement for each leg read
\begin{equation}
\label{def}
 \langle {\bf r}_j \rangle = 0, \quad  \langle {\bf r}_j^2 \rangle= 4t, \quad j=1,\ldots,L
\end{equation}
where we have assumed that initially all legs are at the origin. (We continue to assume that the spider moves on the square lattice; generally on the cubic lattice in $d$ dimensions the amplitude on the right-hand side of  \eqref{def} is given by $2d$.) Using \eqref{def} one computes the variance of the center of mass \eqref{centermass} of the spider
\begin{equation*}
\langle {\bf R}^2 \rangle =L^{-2}\sum_{j=1}^L \langle {\bf r}_j^2 \rangle = L^{-1}\,4t
\end{equation*}
implying that the diffusion coefficient $D(L)$ of the spider with $L$ non-interacting legs  (each performing the random walk with diffusion coefficient $D\equiv 1$) is
\begin{equation}
\label{DL_lim}
D(L) = \frac{1}{L}
\end{equation}
This result is remarkably universal: It is valid in any dimension and it also does not depend on the lattice, the only requirement is the absence of bias. 

\subsection{Interacting Legs}

We now return to our original spiders which move on a lattice and obey two rules:
(i) Two legs cannot be attached to the same site; and (ii) the distance between any two legs does not exceed $S$. In the $S\to\infty$ limit the legs are asymptotically non-interacting. Therefore Eq.~\eqref{DL_lim} implies that $D(L,S=\infty)=1/L$. 

The non-trivial task is to compute the leading correction which describes the deviation from \eqref{DL_lim} in the situation when the maximal span $S$ is large, but finite. To guess the $S$ dependence of the leading correction let us look at known {\em exact} results for spiders moving on the one-dimensional lattice. For the bipedal spider the diffusion coefficient is given by \cite{spider1d}
\begin{equation}
\label{D2_S}
D(2,S) = \frac{1}{2}\left(1-\frac{1}{S}\right)
\end{equation}
More generally in one dimension for multi-leg spiders with nearest-neighbor hopping and the constraint on the maximal span (spiders with global constraint in terminology of Ref.~\cite{spider1d}), the diffusion coefficient reads \cite{spider1d} 
\begin{equation}
\label{DL_S}
D(L,S) = \frac{1}{L}\left(1-\frac{L-1}{S}\right)
\end{equation}
This formula is valid for $L=1$ (the random walk) and all $L\geq 2, S\geq L-1$. 

Equations \eqref{D2_S}--\eqref{DL_S} show that at least in one dimension the leading correction has the $S^{-1}$ dependence on the maximal span. We now demonstrate the universality of this behavior by analyzing the bipedal spider in one dimension with a more complicated gate. Namely, let us assume two types of hops: the nearest-neighbor $\pm 1$ hops occur with rate $p$ each, and the next-nearest-neighbor $\pm 2$ hops occur with rate $q$ each. For a random walk with such gate, the diffusion coefficient is $D=p+4q$, and since we always normalize this diffusion coefficient of the random walk to unity, we set $1=p+4q$. Using methods of Ref.~\cite{spider1d},  one obtains
\begin{equation}
\label{D2_Sp}
D(2,S;p) = \frac{1}{2}\left(1-\frac{2-p}{S}\right)
\end{equation}
showing that the $S^{-1}$ behavior is universal, namely it is insensitive to the details of the gate.  

In two dimensions all known results also agree with the $S^{-1}$ behavior of the leading correction \cite{ell_S}. The exact diffusion coefficient for the von Neumann bipedal spider \eqref{D-bi-S} is in perfect agreement (no other corrections). For the Euclidean bipedal spider the behavior is more complicated [see \eqref{D-vs-L}]; the asymptotic behavior is very simple,
\begin{equation}
\label{D2_S-E}
 D(2,\infty) - D(2,S) \simeq \frac{1}{\pi S}\,,
\end{equation}
and it agrees with the $S^{-1}$ asymptotic. Thus the exact results \eqref{D-bi-S}, \eqref{D2_S}--\eqref{D2_Sp}, and the asymptotic \eqref{D2_S-E} suggest the following conjectural asymptotic behavior 
\begin{equation}
\label{DL-S-d}
 D(L,\infty) - D(L,S) \simeq \frac{A_d(L)}{S}\quad \text{as}\quad S\to\infty
\end{equation}
Intuitively, the $1/S$ correction stems from the ratio of spider-leg configurations where the constraint is relevant (the distance of two or more legs is $S$) to the total number of such configurations. The total number of configurations is essentially a volume of a domain of characteristic size $S$, while the number of configurations for which the constraint is relevant is the surface area of that domain, so the ratio is indeed  proportional to $1/S$. 

We have presented evidence in favor of the conjectural behavior \eqref{DL-S-d} in one and two dimensions, but it is probably valid in arbitrary dimension $d$. The limiting diffusion coefficient $D(L,\infty)=1/L$ is universal, while the amplitude $A_d(L)$ in the sub-leading term depends not only on the number of legs, but also on the spatial dimension $d$, and on the details of the gate. In all known examples the amplitude $A_d(L)$ is positive, e.g. according to \eqref{DL_S} the amplitude is $A_1(L)=1-L^{-1}$ in one dimension. The positivity of $A_d(L)$ is physically evident (the constraint on the maximal span makes spiders less motile), although it is not clear how to prove this positivity.  Finally we note that simulation results for diffusion coefficients presented in Table~\ref{Dtab} are in surprisingly good agreement with the conjectural asymptotic \eqref{DL-S-d}, namely the quantity $S[D(L,\infty) - D(L,S)]$ already changes very little when $S$ increases from 2 to 3.

\section{Non-Markovian effects}
\label{memory}

In this section we continue to assume that the re-attachment of a leg is instantaneous. In contrast to Secs.~\ref{von} and \ref{bi}, however, we take into account the effects of memory associated with previous visits of the legs. These effects are unavoidable in most experimental realizations --- the first time a leg visits a site (an un-cleaved substrate), the leg hops from this site only after it has cleaved it into a product. (The product is unaffected by future visits, so we only need to know if the site has been visited in the past or not.) Thus the motion of the synthetic molecular motor, the molecular spider, irreversibly changes the environment making the problem non-Markovian. Intriguingly, one natural molecular motor, a special protein called collagenase which moves along collagen fibrils, exhibits even stronger irreversible effect on its one-dimensional track and undergoes a biased diffusion \cite{sci,AK,Kol}.

Our main interest is the leading order behavior of two quantities: the mean square position $\langle{\bf R}^2 \rangle \sim 4Dt$, i.e. the diffusion coefficient, and the mean number of visited sites $\langle N \rangle \sim At/\ln t$. We start with the one-leg spider, the random walk, where the effect of memory is asymptotically negligible, yet the corrections to the leading behaviors are qualitatively similar to those which arise for multi-legs spiders. 

The results reported in this section are mostly numerical. In two dimensions, all simulations are made for von Neumann spiders. The convergence to the true asymptotic behaviors is slow, especially in two dimensions where it is logarithmically slow. Therefore to extract accurate numerical predictions for the diffusion coefficient $D$ and the amplitude $A$ we need to know functional forms of correction terms. We make use [in Eqs.~\eqref{wDmemcorr}--\eqref{R_memcorr}] of correction terms which provide good fit to the data. The functional forms of these leading correction terms are still lacking theoretical justification. 

\subsection{Random walk, $L=1$}

We consider the random walk which hops symmetrically to nearest-neighbor sites and changes substrates into products. As in Ref.~\cite{memo}, the change of the substrates to the products (which occurs after the leg first visits the site) is modeled by postulating that the hopping rate to each neighbor is equal to 1 for the site which has not been visited before (the substrate), and to $r$ for the site which has been visited in the past. Thus the hopping rate is determined by the state of the issuing site, but not by the state of the target site. Mathematically, the parameter $r$ can be an arbitrary positive number, $r>0$; in experiments, the detachment from the product is easier, so $0<r<1$ as we shall assume in the following.

According to simulations, the memory has no effect on a random walk in the leading order. Independently of the value of $r$, we obtained $D=1$ for the diffusion coefficient, and $A=4\pi$ for the amplitude. The same universality was numerically observed in one dimension \cite{memo}; for the amplitude $A$, the lack of dependence on $r$ was established analytically \cite{memo}.

To extract the leading order behavior from the simulation data one has to investigate higher order corrections as well. Our numerical simulations indicate that the mean square position has the following correction terms in the presence of memory
\begin{equation}
\label{wDmemcorr}
 \langle{\bf R}^2 \rangle \approx \left\{
 \begin{tabular}{ll}
 $2Dt \left[1 - a_1(r)/\sqrt{t}\right]$ ~~~ &: $d=1$\\
 $4Dt \left[1- a_2(r)/\ln t \right]$ &: $d=2$
 \end{tabular}
 \right.
\end{equation}
while there are no corrections at all in the absence of memory $r=1$, that is, $a_d(1)=0$. The large corrections in two dimensions are especially important in determining the diffusion coefficient. Fitting our data to \eqref{wDmemcorr} we got $D=1$ both in one and two dimensions. The amplitudes of the correction terms are positive, $a_d(r)>0$ when $0<r<1$, and monotonically decreasing functions of $r$.  Therefore the memory slows down the random walk in the next to leading order. This behavior is understandable, as the random walk slows down at newly visited sites. 

For the average number of sites $\langle N \rangle$ visited during the time interval $(0,t)$, the corrections to the leading asymptotic are also more important (since they vanish much more slowly with time) in two dimensions. According to our simulations, 
\begin{equation}
\label{wAmemcorr}
 \langle N \rangle \approx \left\{
 \begin{tabular}{ll}
 $A_1(r)\,\sqrt t \left[1-c_1(r)/\sqrt t \right]$ & : $d=1$\\
 $\frac{A_2(r)\,t}{\ln t} \left[1- c_2(r)/\ln t\right] ~~$ &: $d=2$
 \end{tabular}
 \right.
\end{equation}
The correction amplitudes are again positive, $c_d(r)>0$ when $0<r<1$, and monotonically decreasing functions of $r$.  Fitting our data to \eqref{wAmemcorr} we found that the amplitudes $A_d(r)$ do not depend on $r$ for the random walk. (Equation \eqref{wAmemcorr} also describes the average number of sites visited by a spider, and in that situation the amplitudes $A_d(r)$ do depend on $r$.)
Note that in two dimensions we found the same type of asymptotic behavior both with and without memory,  see Eq.~\eqref{different-spider}, and for an arbitrary number of legs.

\subsection{Multi-pedal molecular spiders, $L>1$}

In order to investigate the effects of memory we numerically studied the two simplest spiders: the bipedal $L=2$, $S=1$ spider, and the tripod $L=3$, $S=2$.

We have seen in Sec.~\ref{von} that when memory effects are ignored, the corrections to the diffusion coefficient $D$ quickly decay with time \eqref{Dnomemcorr}. This feature  makes $D$ easy to measure. In the presence of memory, however, a much slower decaying correction term appears in two dimensions \begin{equation}
\label{R_memcorr}
 \langle{\bf R}^2 \rangle \approx \left\{
 \begin{tabular}{ll}
 $2D_1(r)t \left[1 - a_1(r)/\sqrt{t}\right]$ ~~~ &: $d=1$\\
 $4D_2(r)t \left[1 - a_2(r)/(\ln t)^2\right]$ &: $d=2$
 \end{tabular}
 \right.
\end{equation}
In one dimension, the correction also decays slower with time in the presence of memory, yet it remains algebraic; the qualitative behavior is similar to the random walk with memory [see \eqref{wDmemcorr}]. The $(\ln t)^{-2}$ type correction observed in two dimensions is unusual; it is rather slow of course, yet it is faster than for the random walk. 

\begin{figure}
\centering
\includegraphics[scale=0.7]{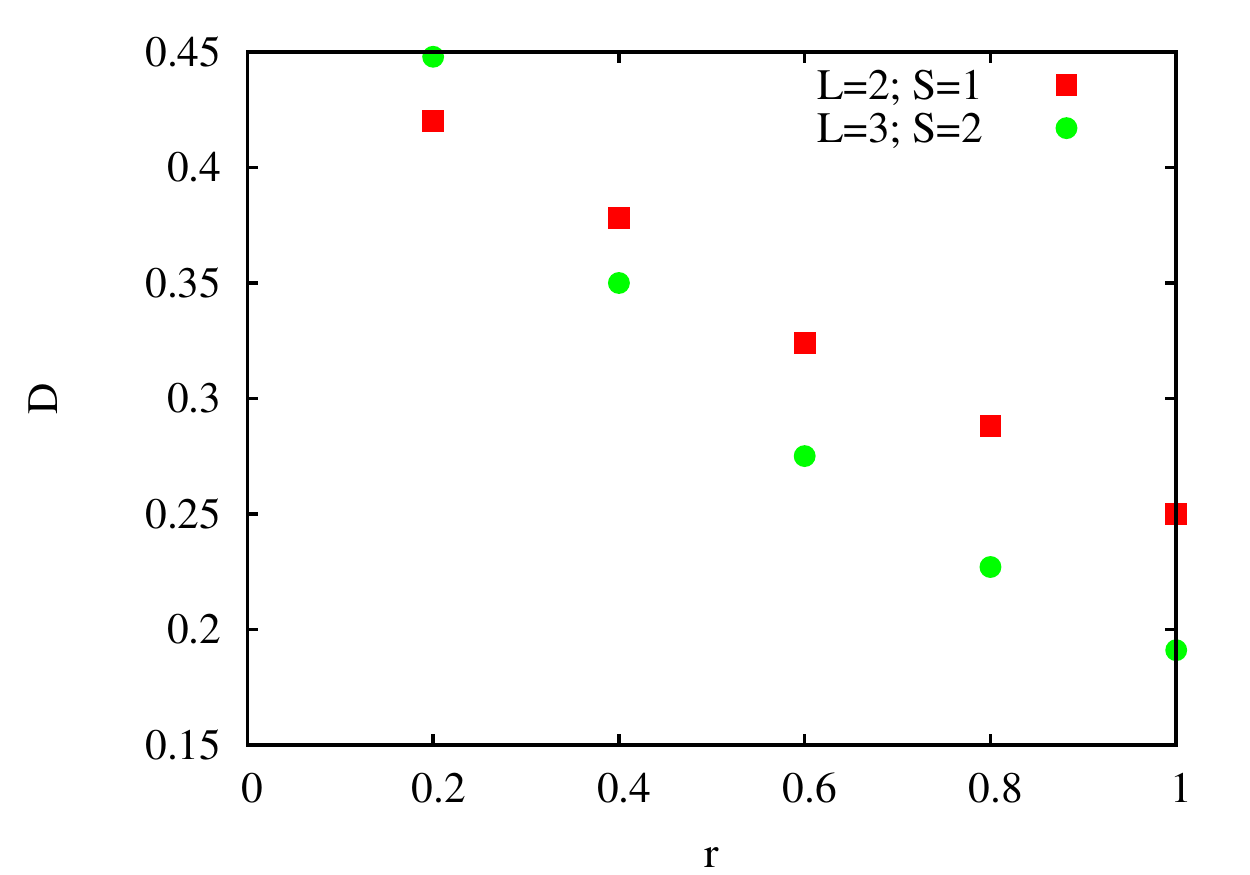}
\caption{Diffusion coefficient $D$ describing the asymptotic behavior of the mean square position of a spider $\langle {\mathbf R}^2 \rangle \sim 4Dt$ as a function of the memory parameter $r$.  Simulation data are shown for the simplest bipedal spider with $S=1$, and for the simplest tripod with $S=2$. The no memory case corresponds to $r=1$.}
\label{D2memoD}
\end{figure}

\begin{figure}
\centering
\includegraphics[scale=0.7]{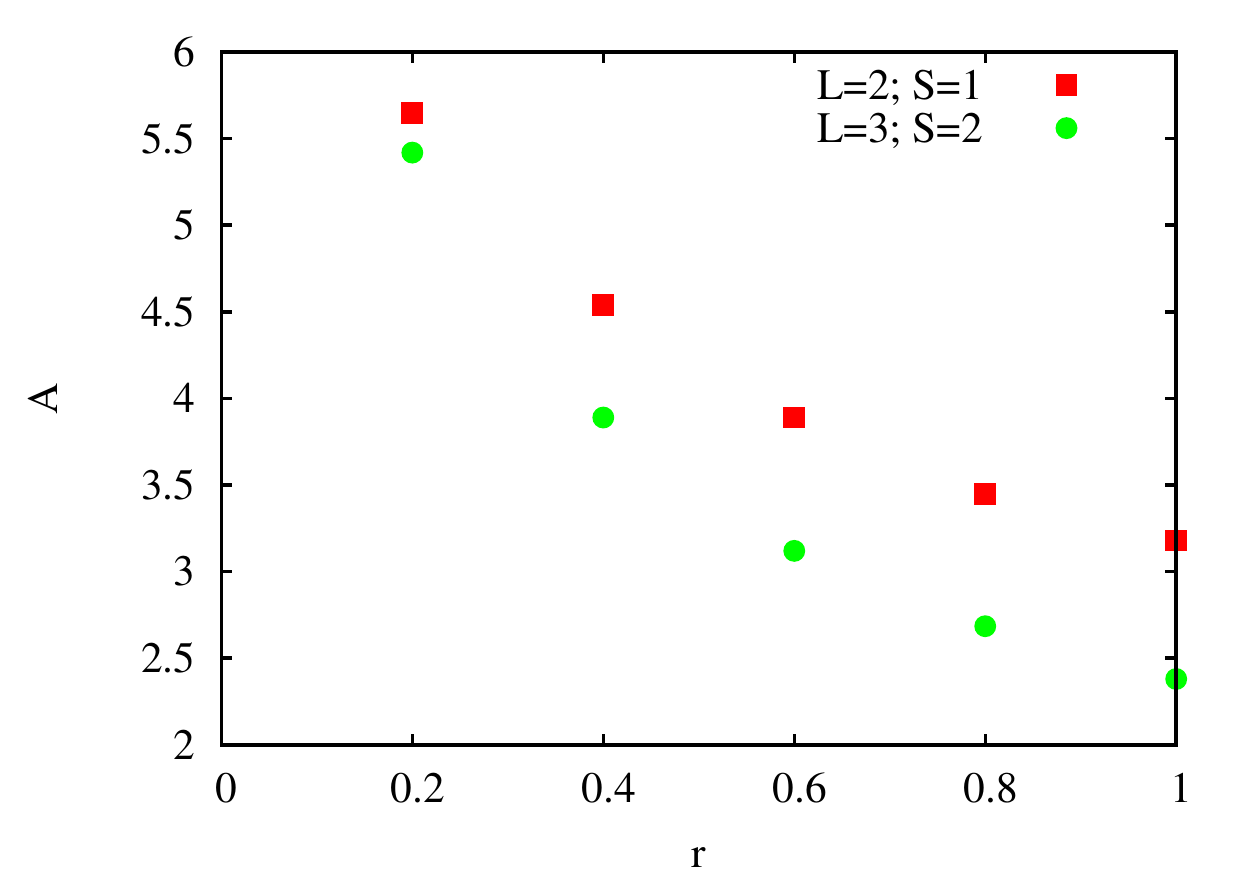}
\caption{Amplitude $A$ describing the asymptotic behavior of the number of visited sites as $\langle N \rangle \sim At/\ln t$ as a function of the memory parameter $r$. Simulation data are shown for the simplest bipedal spider with $S=1$, and for the simplest tripod with $S=2$.}
\label{D2memoA}
\end{figure}

Using \eqref{R_memcorr} and \eqref{wAmemcorr} we extract the diffusion coefficient $D(r)$ and the amplitude $A(r)$; their dependences on $r$ are displayed \cite{index} in Figs.~\ref{D2memoD} and \ref{D2memoA}. Surprisingly, the slowdown of legs at new sites leads to larger diffusion coefficient and increases the average number of visited sites in the large time limit. The qualitative reason is that the slow leg at new sites keeps the other legs close to newly visited sites, which generates an effective bias toward new sites, and thereby it leads to the increase in the number of visited sites. This effect has already been observed, and quantitatively understood, for one-dimensional spiders \cite{memo}. Although here we talk about the asymptotic behavior, the actual average number of visited sites also becomes larger for smaller values of $r$ after about a few hundred time units.

An interesting new feature of  two-dimensional spiders is that the tripod is somewhat more sensitive to the slowdown at new sites than the bipedal spider, as $D$ changes more rapidly with changing $r$ (see Fig.~\ref{D2memoD}). Another remark is that, as one can see in Figs.~\ref{D2memoD} \& \ref{D2memoA}, with large enough slowdown ($r\lessapprox 0.6$) the tripod becomes faster than the bipedal spider without memory ($r=1$). 

An intriguing feature of these two-dimensional spiders is that for sufficiently small $r$ tripods become more motile than bipedal spiders, see in Fig.~\ref{D2memoD}. A possible reason for this effect is that the spider with more legs stick to the domain of new sites more efficiently.

\section{Unbinding of spiders}
\label{binding}

All our previous analyses have relied on the assumption that the re-attachment of a leg is instantaneous, so the process is controlled by detachment. This implies that spiders remain fully attached and never leave the surface.  If the re-attachment takes time, the problem becomes more complicated but not necessarily intractable --- for molecular motors \cite{motors}, for instance, some analyses allow complete detachment (unbinding) from cytoskeletal filaments \cite{det-frey,det-evans,KL,KL1,KL2,traffic,mauro}. 

In this section we treat a more restricted problem, namely we compute the probability that a spider remains attached. We disregard memory and lattice effects and focus on the attachment-detachment process. 

Consider a spider with $L$ legs. Let $\Pi_n(t)$ be the probability that $n$ of its legs are attached. To simplify notation we set the detachment rate equal to unity; we shall disregard the difference between the substrate and the product. We denote the attachment rate by $\lambda$. The probabilities $\Pi_n(t)$ with $n=1,\ldots,L$ obey
\begin{eqnarray}
\label{Pt}
\frac{d\Pi_n}{dt} &=& (n+1)\Pi_{n+1}-n\Pi_n\nonumber\\
                             &-&\lambda(L-n)\Pi_n + \lambda(L-n+1)\Pi_{n-1}
\end{eqnarray}
where the terms on the right-hand side in the top line account for detachment and the terms in the bottom line describe re-attachment. Equation \eqref{Pt} remains applicable to extreme probabilities $\Pi_1$ and $\Pi_L$ if we set
\begin{equation}
\label{extremes}
\Pi_0(t)\equiv 0, \quad \Pi_{L+1}(t)\equiv 0
\end{equation}
The latter relation is obvious (by definition, the spider has $L$ legs); the former relation is actually the assumption that if all legs are detached, the spider leaves the surface and never re-attaches to it. The initial condition is
\begin{equation}
\label{init}
\Pi_n(0)=\delta_{n,L}
\end{equation}
if we imagine that the spider is initially fully attached. 

An analysis of the initial-boundary value problem \eqref{Pt}--\eqref{init} is rather straightforward, so we only present one asymptotic result which is valid in the most interesting limit of quick re-attachment, $\lambda\gg 1$. In this limit we found that, for any $L$, the probability $\Pi(t) = \sum_{n=1}^L \Pi_n(t)$ that the spider remains attached at time $t$ is exponential:
\begin{equation}
\label{L-leg}
\Pi=\exp(-Lt/\lambda^{L-1})
\end{equation}
Equation \eqref{L-leg} shows that the mean time for the spider to detach is
\begin{equation}
\label{t-av}
\langle t_L\rangle=\frac{\lambda^{L-1}}{L}
\end{equation}
in the $\lambda\gg 1$ limit. 

If we were only interested in the mean detachment time, we could have determined it exactly without computing the probabilities $\Pi_n(t)$. Indeed, utilizing an exact solution for the adsorption time for the so-called one-step process \cite{van}, one finds an exact expression
\begin{equation}
\label{t-av-sol}
\langle t_L\rangle = \sum_{a=1}^{L} \lambda^{L-a} \sum_{b=1}^{a} 
 \frac{(L-b)!(b-1)!}{(a-b)!(L-a+b)!}
\end{equation}
which is a polynomial in $\lambda$. Keeping only the dominant $a=1$ term we see that \eqref{t-av-sol} reduces to \eqref{t-av}.  Displaying a few more terms we obtain
\begin{equation*}
 \begin{split}
\langle t_L\rangle &= \lambda^{L-1}\frac{1}{L} + \lambda^{L-2} \left[ 1+ \frac{1}{L(L-1)}\right]\\
 &+ \lambda^{L-3}  \left[  \frac{L-1}{2}  + \frac{1}{L-1} + \frac{2}{L(L-1)(L-2)}\right]+\ldots
 \end{split}
\end{equation*}

The above analysis is ``zero-dimensional'' as we ignored the lattice. For instance, consider a tripod in the configuration $\bullet\bullet\bullet$ and imagine that one of its legs detaches. If an extreme leg detaches, its re-attachment rate is clearly larger than the re-attachment rate of the middle leg. In equations \eqref{Pt} this feature is ignored. Therefore combining detachment with diffusion makes the problem very complicated even in the absence of memory effects. A numerical analysis of models which take into account the possibility of the detachment has been undertaken in Refs.~\cite{zucker,zucker2}. For the bipedal molecular spider, the one-dimensional version of this problem could be tractable as long as the memory effects are ignored.

\section{Summary}
\label{disc}

We investigated the motion of a single molecular spider on the square lattice. The limit when the motion of the spider does not affect the environment is tractable for bipedal spiders, while spiders with more than two legs remain very challenging for analytical work. For bipedal spiders we computed the diffusion coefficient with an arbitrary maximal span between the legs. Generally, the increase in the maximal span leads to the increase in the diffusion coefficient. This phenomenon strictly holds for one type of bipedal spiders (von Neumann), while for Euclidean bipedal spiders an increase of the span can sometimes, unexpectedly, decrease the diffusion coefficient.  

We explored the behavior of von Neumann spiders with more than two legs by means of numerical simulations. In general we found that the increase of the number of legs makes spiders less motile. The increase of the maximal span $S$ (maximal allowed distance between the legs), on the other hand, makes spiders more motile. In the infinite span limit the diffusion coefficient is reciprocal to the number of legs, $D(L,S=\infty)=1/L$. We argued that the leading large $S$ correction to the diffusion coefficient is proportional to $1/S$. We also considered the effect of spiders completely unbinding from the substrate, and we found that the time it takes grows exponentially with the number of legs. The reason is that a spider unbinds only if all of its legs simultaneously unbind.

In experimental realizations, the legs usually convert substrates into shorter products that have a lower affinity for the legs. Assuming that the substrate turns into the product after the first visit of a leg, we investigated this non-Markovian problem numerically. We showed that the long-time behavior is diffusive in character. More precisely, we demonstrated that the mean-square displacement grows as $t$ and the total number of distinct visited sites grows as $t/\ln t$. The amplitudes are affected by memory. Furthermore, the non-Markovian nature of the problem leads to very large sub-leading corrections. For instance, the relative magnitude of the sub-leading correction to the mean-square displacement decays with time in a very slow $(\ln t)^{-2}$ manner. 

The most surprising influence of memory is that the more spiders are slowed down on unvisited sites, the more motile spiders become. An explanation of this very counter-intuitive behavior is that the ``stickiness" to unvisited sites generates an effective bias toward unvisited sites, which results in the increase of the visited area. For example, although without memory an $L=3, S=2$ spider is slower than an $L=2, S=1$ spider, with the memory effect the three leg spider can visit more sites on average.

\acknowledgments{We thank Darko Stefanovic and Milan Stojanovic for helpful discussions and suggestions. PLK acknowledges financial support  from NSF Grant No. CCF-0829541.}

\bigskip

\appendix

\section{EUCLIDEAN BIPEDAL SPIDERS}
\label{bipedal}

For small $\ell$, we can count various configurations and compute their rates
by hand. Here we record what happens when $\ell$ varies up to $\sqrt{50}$.
Configurations that arise when $\ell$ varies up to $\sqrt{20}$ are depicted below. 
\begin{figure}[ht]
\vspace*{0.cm}
\includegraphics*[width=0.45\textwidth]{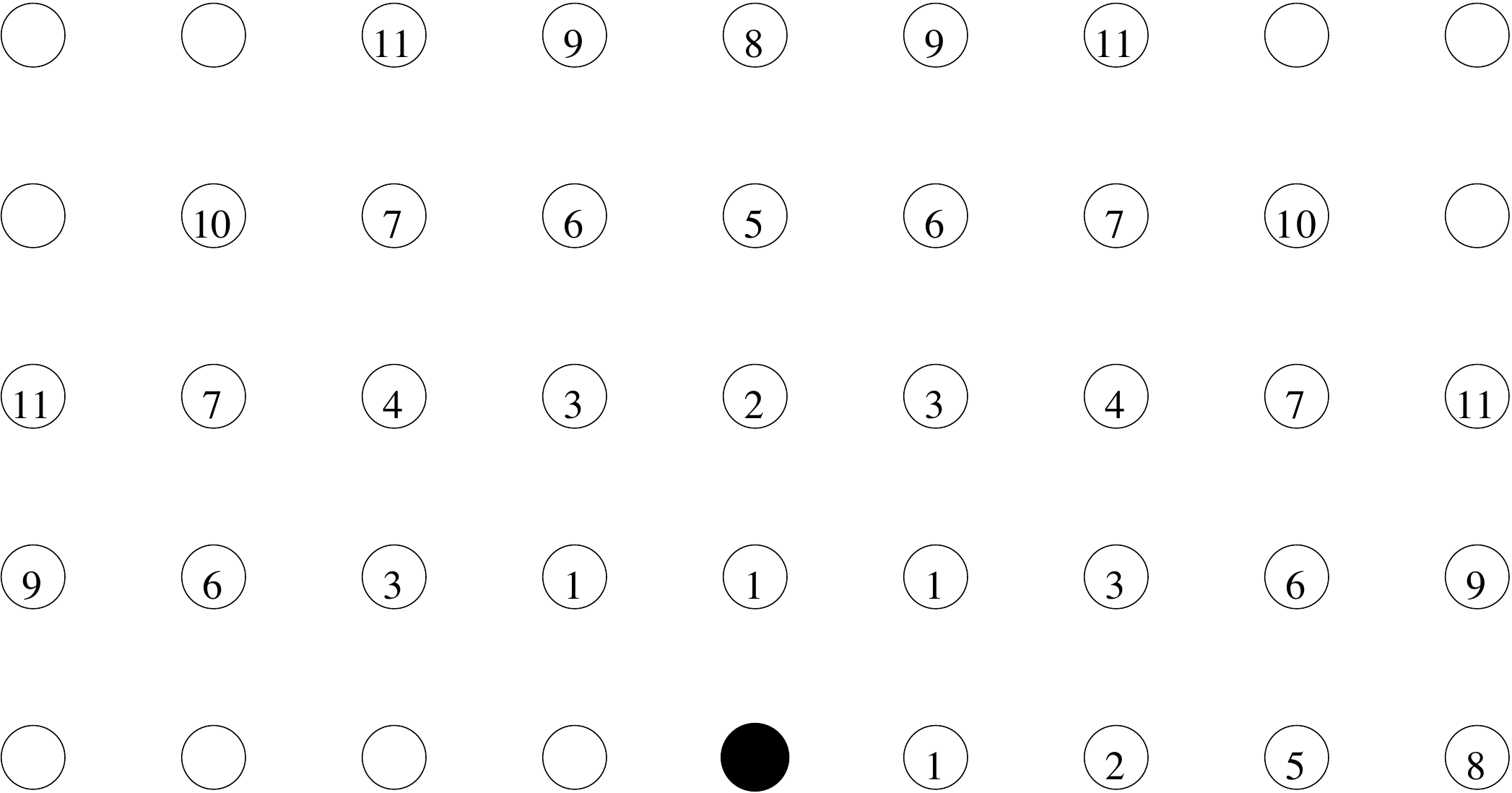}
 \caption{Non-isomorphic configurations for $\ell=\sqrt{20}$.  Configurations
   share the center site (filled disk) while another leg is in a labeled
  disk. The label counts the `bifurcation' event when such configurations
  first appear (the first bifurcation is identified with $\ell=\sqrt{2}$).
   For instance, the simplest knight configurations appear in the third
   bifurcation event (when $\ell$ passes through $\sqrt{5}$) and hence the
   corresponding label is 3.}
\label{lattice}
\end{figure}

\subsection{$\sqrt{2}\leq \ell<2$} 

There are four different configurations.  The
rates are $\omega_1=\omega_2=\omega_3=\omega_4=4$ and hence (\ref{compD2}) gives $D=1/4$.

\subsection{$2\leq \ell<\sqrt{5}$}

Two additional configurations are $\bullet\circ\bullet$ and its vertical
cousin, so that $\mathcal{C}=6$.  We have
\begin{equation*}
\omega(\bullet\bullet)=6,\qquad \omega(\bullet\circ\bullet)=2,\qquad \omega_{\rm diag}=8
\end{equation*}
and therefore
\begin{equation}
\label{D_6}
D=\frac{1}{16}\,\frac{1}{6}\,[6+2+4]\times 2=\frac{1}{4}
\end{equation}
Surprisingly, the diffusion coefficient is the same as for the previous
two-leg spider.

\subsection{$\sqrt{5}\leq \ell<2\sqrt{2}$}

Four more configurations become possible: $\mathcal{C}=10$.  The
rates of these new knight configurations (Fig.~\ref{lattice}) are $\omega_{\rm
  knight}=4$.  The rates of the previous configurations are
\begin{equation}
\label{horizontal} 
\omega(\bullet\bullet)=\omega(\bullet\circ\bullet)=6
\end{equation}
while the diagonal configurations reach the maximal possible rate $\omega_{\rm diag}=8$. 
Thus
\begin{equation*}
D=\frac{1}{16}\,\frac{1}{10}\,[6\times 4+8\times 2+4\times 4]=\frac{7}{20}
\end{equation*}

\subsection{$2\sqrt{2}\leq \ell<3$}

Two new diagonal configurations (Fig.~\ref{lattice}) of length $2\sqrt{2}$ become possible; overall $\mathcal{C}=12$.  The rates remain the same [Eq.~(\ref{horizontal})], for horizontal and vertical configurations of both kinds, and for short diagonal configurations ($\omega_{\rm diag}=8$).  The rates
increase for knight configurations: $\omega_{\rm knight}=6$. Finally for the new
diagonal configurations (Fig.~\ref{lattice}) the rates are $\omega_{\rm
  diag}^{(2)}=4$.  Thus
\begin{equation*}
D=\frac{1}{16}\,\frac{1}{12}\,[6\times 4+8\times 2+6\times 4 + 4\times 2]=\frac{3}{8}
\end{equation*}
Note that as the number of configurations from 6 to 10 to 12, the diffusion coefficient 
also gets larger. 

\subsection{$3\leq \ell<\sqrt{10}$}

The rates of horizontal configurations are
\begin{equation}
\label{hor} 
\omega(\bullet\bullet)=6,\quad  \omega(\bullet\circ\bullet)=8,\quad  \omega(\bullet\circ\circ\,\bullet)=2
\end{equation}
and similarly for the vertical once. The other rates are
\begin{equation}
\label{diag} 
\omega_{\rm diag}=8,\quad  \omega_{\rm diag}^{(2)}=4,\quad  \omega_{\rm knight}=6
\end{equation}
The diffusion coefficient $D=5/14$ is {\em smaller}
than the diffusion coefficient $D=3/8$ that characterizes spiders with
$2\sqrt{2}\leq \ell<3$.

\subsection{$\sqrt{10}\leq \ell<\sqrt{13}$}

Four new knight configurations obtained by hopping one lattice spacing in one direction and three lattice spacings in orthogonal direction arise. The rates which change compared to rates (\ref{hor})--(\ref{diag}) are
\begin{equation*}
\omega(\bullet\circ\circ\,\bullet)=6,\quad \omega_{\rm knight}=8,\quad \omega_{\rm knight}^{(2)}=4
\end{equation*}
The diffusion coefficient is $D=7/18$.

\subsection{$\sqrt{13}\leq \ell<4$}

Four new knight configurations obtained by hopping two lattice spacings in one direction and three lattice spacings in orthogonal direction arise.  The rates which change are
\begin{equation*}
\omega_{\rm diag}^{(2)}=8,\quad \omega_{\rm knight}^{(2)}=6,\quad \omega_{\rm knight}^{(3)}=4
\end{equation*}
The diffusion coefficient is $D=9/22$.

\subsection{Large $\ell$}

We now outline following bifurcations. We classify various configurations into linear, diagonal, and knight. Each linear configuration is either horizontal or vertical. There are also two kinds of diagonal
configurations while knight configurations have four different types. The rates of linear configurations and diagonal configurations [by $(d,d)$ we denote the diagonal configuration of length $d\sqrt{2}$] change with their size. Some of these rates are collected in Table \ref{diagtab} (we present results in the window $5\leq \ell\leq 7$; overall we analyzed $\ell\leq 11$).

\begin{table}
\begin{tabular}{|r|r|r|r|r|r|r|r|r|r|r|r|r|}
\hline
$\ell$&1&2&3&4&5&6&7&(1,1)&(2,2)&(3,3)&(4,4)\\
\hline
5          &6&8&8&8&2&0&0&8&8&8&0\\
$\sqrt{26}$&6&8&8&8&6&0&0&8&8&8&0\\
$\sqrt{29}$&6&8&8&8&6&0&0&8&8&8&0\\
$\sqrt{32}$&6&8&8&8&6&0&0&8&8&8&4\\
$\sqrt{34}$&6&8&8&8&6&0&0&8&8&8&4\\
6          &6&8&8&8&8&2&0&8&8&8&4\\
$\sqrt{37}$&6&8&8&8&8&6&0&8&8&8&4\\
$\sqrt{40}$&6&8&8&8&8&6&0&8&8&8&4\\
$\sqrt{41}$&6&8&8&8&8&6&0&8&8&8&8\\
$\sqrt{45}$&6&8&8&8&8&6&0&8&8&8&8\\
7          &6&8&8&8&8&8&2&8&8&8&8\\
\hline
\end{tabular}
\caption{The rates of the linear and diagonal configurations for Euclidean spiders in the $5\leq \ell\leq 7$ window.}
\label{diagtab}
\end{table}

In Table \ref{knighttab} we summarize how the rates of the knight configurations ($nm$ denotes the knight
configuration obtained by hopping $n$ lattice spacings in one direction and
$m$ lattice spacings in another) vary with $\ell$. 

\begin{table}
\begin{tabular}{|r|r|r|r|r|r|r|r|r|r|r|r|r|r|r|r|r|}
\hline
$\ell$&21&31&32&41&42&43&51&52&53&54&61&62&63\\
\hline
5          &8&8&8&6&6&4&0&0&0&0&0&0&0\\
$\sqrt{26}$&8&8&8&8&6&4&4&0&0&0&0&0&0\\
$\sqrt{29}$&8&8&8&8&8&4&6&4&0&0&0&0&0\\
$\sqrt{32}$&8&8&8&8&8&6&6&4&0&0&0&0&0\\
$\sqrt{34}$&8&8&8&8&8&8&6&6&4&0&0&0&0\\
6          &8&8&8&8&8&8&6&6&4&0&0&0&0\\
$\sqrt{37}$&8&8&8&8&8&8&8&6&4&0&4&0&0\\
$\sqrt{40}$&8&8&8&8&8&8&8&8&4&0&6&4&0\\
$\sqrt{41}$&8&8&8&8&8&8&8&8&6&4&6&4&0\\
$\sqrt{45}$&8&8&8&8&8&8&8&8&8&4&6&6&4\\
7          &8&8&8&8&8&8&8&8&8&4&6&6&4\\
\hline
\end{tabular}
\caption{The rates of the knight configurations for Euclidean spiders in the $5\leq \ell\leq 7$ window.}
\label{knighttab}
\end{table}

We now give some explanations. The behavior of the quantity $\Phi\equiv (1/2-D)\,\mathcal{C}$ (see Table \ref{euctab}) is easy to understand. First, using Eq.~\eqref{compD2} we can re-write $\Phi$ as 
\begin{equation}
\label{Phi} 
\Phi=\frac{1}{2}\,\mathcal{C} - \frac{1}{16}\,\mathcal{R}\,,\qquad
\mathcal{R}=\sum_{j=1}^{\mathcal{C}} r_j
\end{equation}
Now let us examine the increment of $\Phi$ that occurs when new
configurations are born. A direct counting gives 
\begin{subequations}
\begin{align}
&(\Delta \mathcal{C}, \Delta \mathcal{R})_{\rm diag}=(2,16)
\label{diagonal}\\
&(\Delta \mathcal{C}, \Delta \mathcal{R})_{\rm linear}=(2,8)
\label{linear}\\
&(\Delta \mathcal{C}, \Delta \mathcal{R})_{\rm knight}=(4,32)
\label{knight}
\end{align}
\end{subequations}
Plugging (\ref{linear})--(\ref{knight}) into (\ref{Phi}) we find that 
\begin{equation}
\label{Phi-various} 
(\Delta\Phi)_{\rm linear}=\frac{1}{2}\,,\qquad 
(\Delta\Phi)_{\rm diag}=(\Delta\Phi)_{\rm knight}=0
\end{equation}
Thus $\Phi$ increases by 1/2 when a pair of linear configurations are born and does not change when configurations of other types are added. This explains why $\Phi$ is half-integer and why the jumps in $\Phi$ occur when $\ell$ passes the integer value. Hence 
$\Phi=(\lfloor\ell\rfloor+1)/2$, where $\lfloor\ell\rfloor$ is the integer part of 
$\ell$, that is, the largest integer not exceeding $\ell$. Plugging this into 
$D=1/2-\Phi/\mathcal{C}$ we arrive at Eq.~\eqref{D-vs-L}.


\begin{thebibliography}{99}
\bibitem{D_convention}
    Throughout this paper we set both the lattice spacing and the hopping rate of a leg from the 
    substrate to unity, so that for a one-leg spider (a random walk) the diffusion coefficient is $D=1$.
    With our conventions, the diffusion coefficient of any spider is dimensionless.
\bibitem{ell_S} 
   In this section we use the same notation $S$ for the maximal span both for von Neumann
   and Euclidean spiders. In Sec.~\ref{bi} and the Appendix we use notation $\ell$ for the 
   maximal span of Euclidean spiders. One reason is that e.g. the simplest motile bipedal 
   von Neumann spider has $S=1$ and it is identical to the simplest motile bipedal Euclidean spider
   which has $\ell=\sqrt{2}$.
\bibitem{index}
   $D_2(r)$ and $A_2(r)$ are actually displayed on Figs.~\ref{D2memoD} and \ref{D2memoA}. 
   In Eqs.~\eqref{wDmemcorr}--\eqref{R_memcorr} we used an index to distinguish between dimensions
   $d=1$ and $d=2$, but throughout the paper we are using the two-dimensional framework, so 
   we suppress the index.
\end{thebibliography}
\end{document}